\documentclass[aip]{revtex4-1}
\usepackage{nicefrac}
\usepackage{SIunits}
\usepackage{xcolor}
\usepackage{graphicx}
\usepackage{amsmath}

\begin{document}

\title{Improvement of the transport properties of a high-mobility electron system by intentional parallel conductance}

\author{S. Peters}
\email[]{peters@phys.ethz.ch}
\author{L. Tiemann}
\author{C. Reichl}
\author{S. F\"alt}
\author{W. Dietsche}
\author{W. Wegscheider}
\affiliation{Solid State Physics Laboratory, ETH Zurich, 8093 Zurich, Switzerland}

\date{\today}

\begin{abstract}
	We present a gating scheme to separate even strong parallel conductance from the magneto-transport signatures and properties of a two-dimensional electron system. By varying the electron density in the parallel conducting layer, we can study the impact of mobile charge carriers in the vicinity of the dopant layer on  the properties of the two-dimensional electron system. It is found that the parallel conducting layer is indeed capable to screen the remote ionized impurity potential fluctuations responsible for the fragility of fractional quantum Hall states. 
\end{abstract}

\pacs{}

\maketitle 

Modern high-mobility heterostructures rely on the modulation doping concept that places the dopants slightly away from the heterojunction or quantum well that harbors the two-dimensional electron system (2DES). When the doping concentration is very high or for a low aluminum content in the Al$_{x}$Ga$_{1-x}$As spacer between doping site and interface, the conduction band edge can drop below the Fermi energy at the doping site, resulting in an additional conducting channel. The Ohmic contacts to the 2DES, which are formed by metal deposition and thermal annealing, will also be electrically connected to this parallel channel, resulting in what is generally known as "parallel conductance". The existence of such parallel conductance is signaled by an increasing resistive background of the longitudinal resistance and a bending of the Hall resistance \cite{Grayson2005}. Masking the interesting quantum phenomena, parallel conductance is considered undesirable. However, the mobile charges in the parallel channel generally have an extremely low mobility so that the parallel channel usually does not exhibit Shubnikov-de Haas oscillations or quantum Hall effects which would be superimposed to the signal from the 2DES. Quite the contrary do samples with a slight parallel conductance often exhibit beautifully developed fractional quantum Hall (FQH) effects. This strong development of FQH states can be traced back to the effective screening from the dopants by the parallel conducting layer at the sites of the dopants \cite{Gamez2013}.

Here, we outline a simple top-gating technology that will allow us to utilize samples displaying parallel conductance on the surface side of a single interface. The parallel conductance is isolated from the Ohmic contacts in the arms of a Hall-bar geometry and can be modulated in the bar area itself. Thus, the detrimental effect of parallel conductance can be eliminated and additional impacts on the 2DES characteristics become accessible. We demonstrate that the existence of a parallel conducting layer (PCL) seems to enhance the development of FQH states via the screening of remote ionized impurities. When structured back gates are implemented\cite{Berl2016}, our method can be generalized for inverted single interfaces or double-sided doped quantum wells. We will explain the technological part using a low-quality single-interface modulation-doped GaAs/AlGaAs structure with three consecutive $\delta$-doping layers \unit{5}{nm} apart which result in strong intrinsic parallel conduction (Sample A). The 2DES is buried \unit{245}{nm} below the surface with a setback distance to the delta doping site of \unit{70}{nm}. To illustrate the screening effect of a PCL on FQH states, a second single-interface structure of higher quality was investigated (Sample B). Sample B is quantum well doped at a setback distance of \unit{70}{nm} from the GaAs/AlGaAs interface that is situated \unit{100}{nm} below the surface. In both samples, the evaporation and annealing of Au/Ge Ohmic contacts short the conducting doping layer and the 2DES.

\begin{figure}
	\includegraphics[scale=0.9]{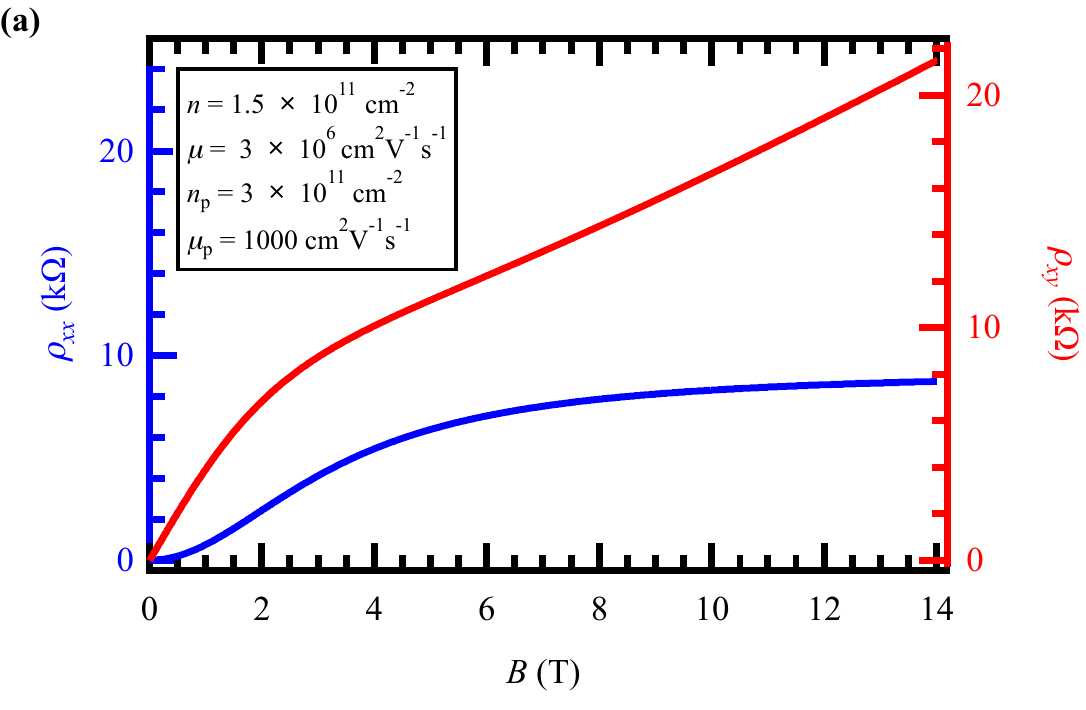}
	\includegraphics[scale=0.9]{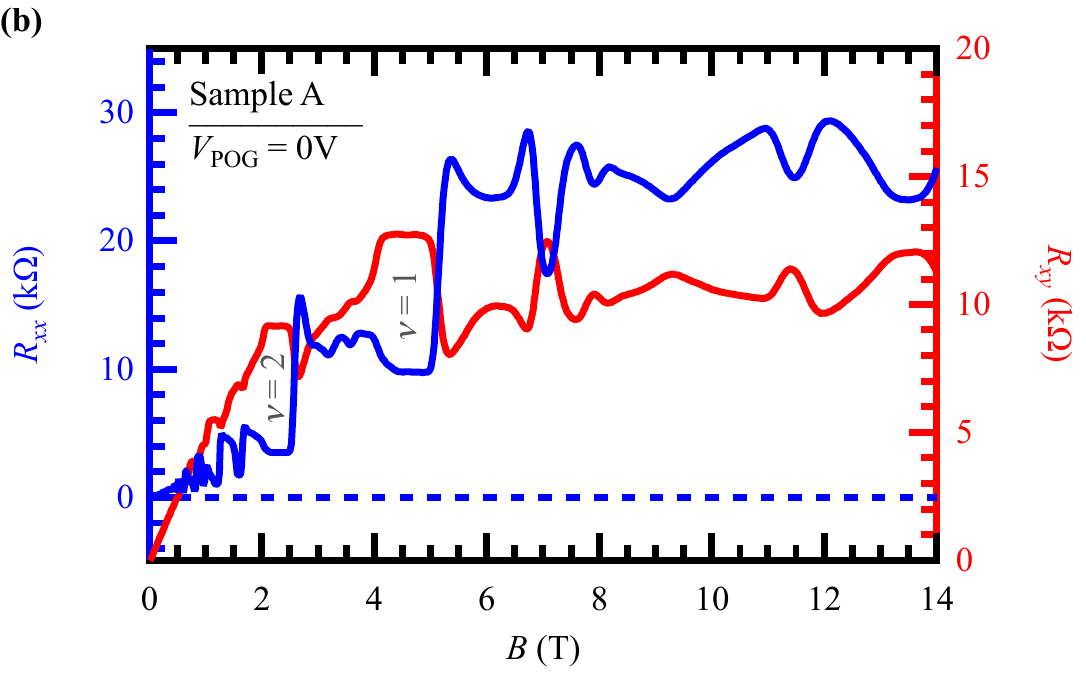}
	\includegraphics[scale=0.9]{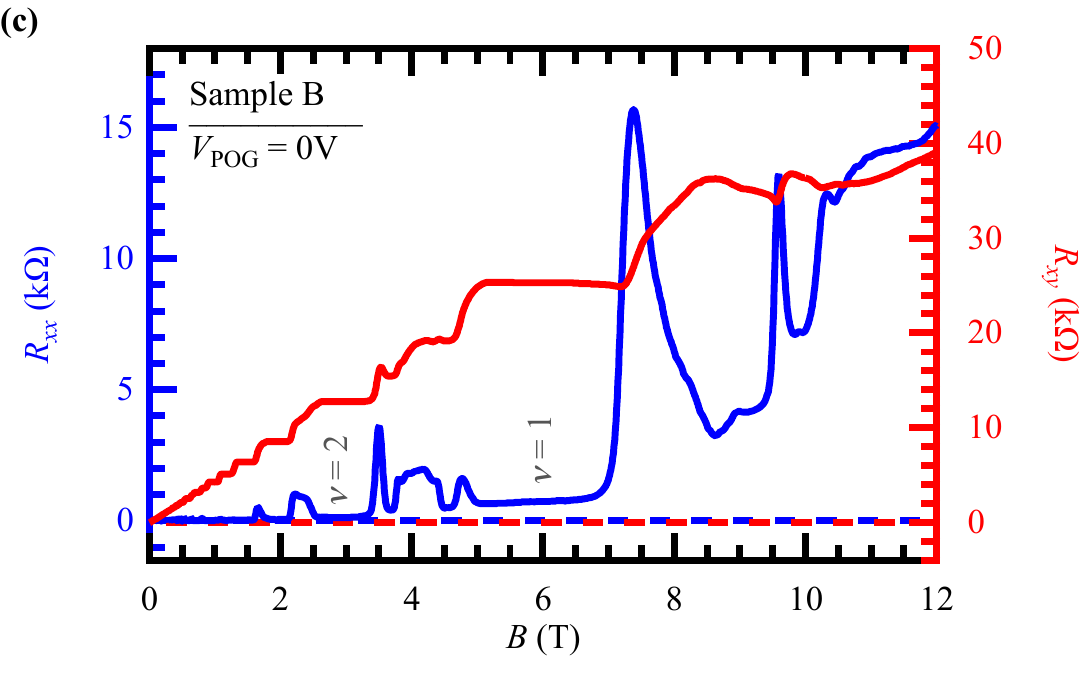}
	\caption{Parallel conduction in the doping layer. {\bf (a)} Profiles of longitudinal ($\rho_{xx}$) (blue) and transverse ($\rho_{xy}$) (red) resistivity calculated within the scope of the Two-band model for parallel conduction, according to Eqs.\;\eqref{rxx} and \eqref{rxy}. {\bf (b)} Longitudinal ($R_{xx}$) (blue) and transverse ($R_{xy}$) (red) resistance as a function of magnetic field $B$ in Sample A and {\bf (c)} Sample B when no gate voltages are applied.}
	\label{fig:PC}
\end{figure}

Figure\;\ref{fig:PC} illustrates the impact of a weak and a strong PCL on the magneto-transport properties. While Sam\-ple B displays an increase in the resistive background of the longitudinal resistance at magnetic fields exceeding \unit{5}{T}, Sample A already exhibits this resistive background at very small $B$-fields. Also the Hall voltage is found to diverge from a linear increase with magnetic field.

Apart from the features related to the quantum Hall effect, these experimental profiles are in agreement with the classical description of parallel conduction through two channels by the so-called Two-band model \cite{Chambers1952, VanHouten1988, Kane1985, Harris1991}. According to this model, the total longitudinal and transverse resistivities arise from the total conductivity via matrix inversion:
\begin{align}
		\rho_{xx} &= \frac{\gamma \gamma_\mathrm{p} \left[ne\mu \gamma_\mathrm{p} + n_\mathrm{p}e\mu_\mathrm{p} \gamma\right]} {\left[ne\mu \gamma_\mathrm{p} + n_\mathrm{p}e\mu_\mathrm{p} \gamma\right]^2 + \left[ne\mu^2B \gamma_\mathrm{p} + n_\mathrm{p}e\mu_\mathrm{p}^2B \gamma\right]^2}\label{rxx}\\
		\rho_{xy} &= \frac{B \gamma \gamma_\mathrm{p} \left[ne\mu^2 \gamma_\mathrm{p} + n_\mathrm{p}e\mu_\mathrm{p}^2 \gamma\right] } {\left[ne\mu \gamma_\mathrm{p} + n_\mathrm{p}e\mu_\mathrm{p} \gamma\right]^2 + \left[ne\mu^2B \gamma_\mathrm{p} + n_\mathrm{p}e\mu_\mathrm{p}^2B \gamma\right]^2}  ,\label{rxy}
\end{align}
with $\gamma=1+\mu^2B^2$ and $\gamma_\mathrm{p}=1+\mu_\mathrm{p}^2B^2$; $n$ and $\mu$ represent the 2DES's density and mobility while $n_\mathrm{p}$ and $\mu_\mathrm{p}$ stand for the respective quantities related to the PCL. These modifications to the field dependence of $\rho_{xx}$ and $\rho_{xy}$ compared to the Drude model \cite{Drude1900a} are depicted in Fig.\;\ref{fig:PC}(a). The measured longitudinal resistivity $\rho_{xx}$ rises as a function of the $B$-field while the Hall curve represented by $\rho_{xy}(B)$ is bent downwards.
At low magnetic fields, the measured transverse Hall resistivity is dominated by the 2DES. The more the mobilities $\mu$ and $\mu_\mathrm{p}$ in both parallel systems differ from each other, the higher is the magnetic field to which the low-field limit, dominated by the 2DES, extends.

\begin{figure}
	\includegraphics[scale=0.9]{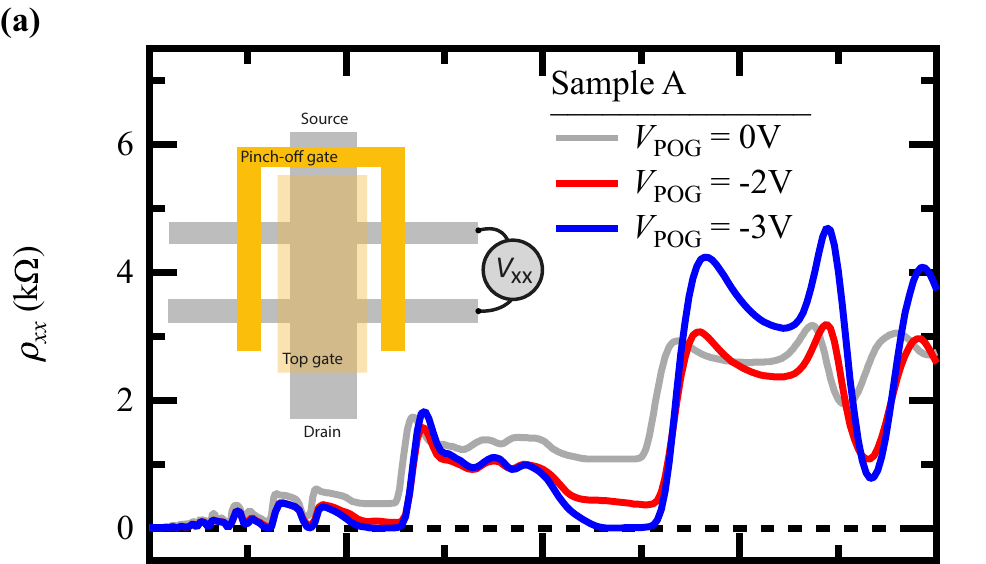}
	\includegraphics[scale=0.9]{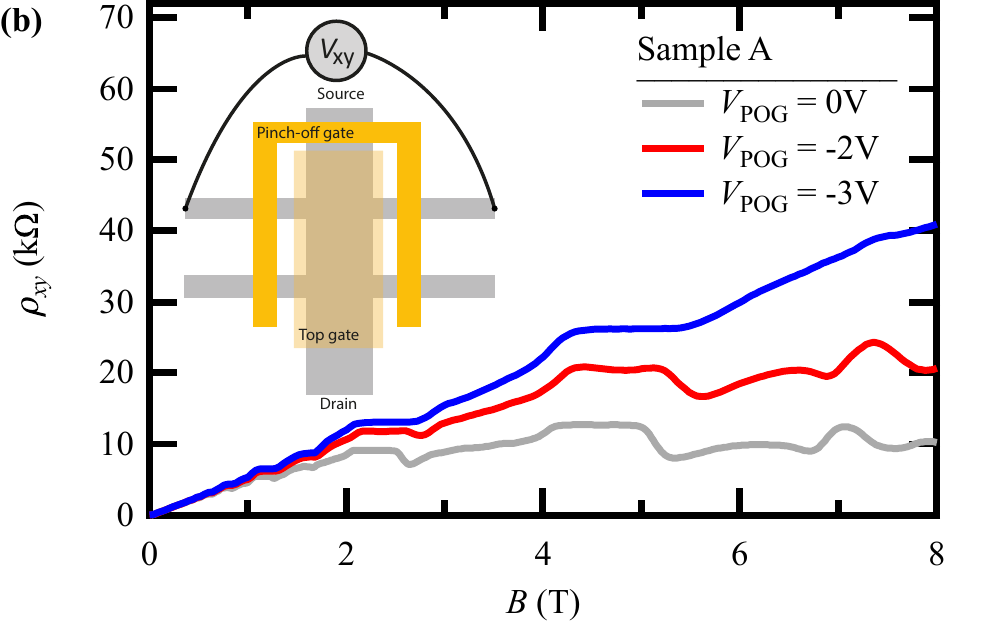}
	\includegraphics[scale=0.9]{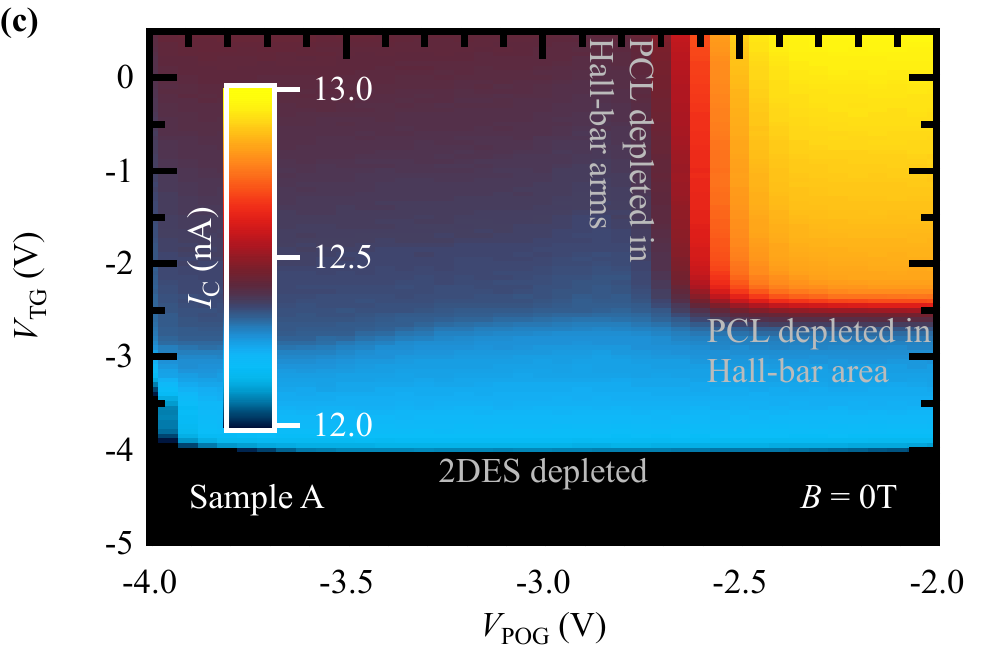}
\caption{{\sl Sample A} ($T\approx\unit{280}{mK}$): {\bf (a)} Longitudinal ($\rho_{xx}$) (blue) and {\bf (b)} transverse ($\rho_{xy}$) (red) resistivity as a function of magnetic field $B$ when the parallel-conducting doping layer is separated from the Ohmic contacts by a negative pinch-off-gate voltage $V_\mathrm{POG}$ ($V_\mathrm{TG}=\unit{0}{V}$). The gray traces correspond to the blue traces in Fig.\;\ref{fig:PC}. {\bf (c)} Capacitance (expressed as the capacitive current $I_\mathrm{C}\propto C_\mathrm{TG}$) of the two-layer system formed by 2DES and PCL with respect to the top gate.}
\label{fig:PO_PCL}
\end{figure}

In order to electrically isolate the 2DES from the PCL, we apply negative voltages to a series of gates placed nearby the Ohmic contacts which will deplete the PCL underneath (see insets of Fig.\;\ref{fig:PO_PCL}). To avoid having the PCL electrically floating, the PCL at the drain contact is not depleted. We emphasize that these “pinch-off gates” (POG) only locally deplete the PCL next to the Ohmic contacts; \emph{they do not eliminate the PCL over the area of the Hall bar}. The control of the PCL density across the Hall bar is achieved via an additional top gate (TG) that covers the entire active Hall-bar region. Varying the top-gate voltage, thus tuning the screening by the PCL, would provide insight into the impact of remote ionized impurities (RIs) on the 2DES.

Figures\;\ref{fig:PO_PCL}(a) and (b) illustrate the dramatic effect on the magneto-transport when the voltage to the POG is made more negative. The pattern of the POG and the TG is sketched in the insets. We observe that the minimum values of $\rho_{xx}$, e.g., the minimum at filling factor $\nu=1$ at $B = \unit{4.9}{T}$, are substantially lowered while at the same time the Hall plateaus at $V_\mathrm{POG}=\unit{-2}{V}$ are lifted up. At $V_\mathrm{POG}=\unit{-3}{V}$, the effect of the PCL is completely eliminated from the magneto-transport data. For pinch-off-gate voltages more negative than $\unit{-3}{V}$, eventually, also the 2DES underneath begins to be depleted. We never operated the POGs in the latter regime.

To determine the exact value of the top gate (and pinch-off) voltage that eliminates the PCL, we employ a capacitive measurement technique. Since the capacitance between top gate and PCL will markedly differ from that between top gate and 2DES, we can easily obtain the top gate voltage that depletes the PCL simply by measuring the capacitance. The capacitance per unit area $C_\mathrm{TG}$ related to the TG and a second conducting layer, the PCL or the 2DES,  is given by
	\begin{equation}\label{eq:CTG}
		\Delta n_\mathrm{layer} = \frac{C_\mathrm{TG}}{e}V_\mathrm{TG} \; ,
	\end{equation}
where $\Delta n_\mathrm{layer}$ refers to the density change in the affected layer (PCL or 2DES) due to the applied top-gate voltage. The pinch-off-gate capacitance per unit area $C_\mathrm{POG}$ is defined accordingly.

We modulate the top-gate voltage $V_\mathrm{TG}$ with AC voltage of $V_\mathrm{TG}^\mathrm{AC}=\unit{10}{mV}$ that has a frequency $f_\mathrm{TG}=\unit{381.3}{Hz}$. We apply the AC-modulated top gate voltage with respect to the drain contact. The capacitive AC current $I_\mathrm{C}$, measured at the drain contact with a standard lock-in technique, is phase shifted by $90^\circ$ relative to $V_\mathrm{TG}^\mathrm{AC}$.  The top-gate capacitance $C_\mathrm{TG}$ is obtained from this off-phase component of the current via
	  \begin{equation}\label{eq:C_IC}
	  	C_\mathrm{TG} = \frac{I_\mathrm{C}}{2\pi f_\mathrm{TG} \cdot V_\mathrm{TG}^\mathrm{AC} \cdot A_\mathrm{TG}} \; ,
	  \end{equation}
with the top-gate area $A_\mathrm{TG}$, $V_\mathrm{TG}^\mathrm{AC}$, and $f_\mathrm{TG}$ being constant.

Figure\;\ref{fig:PO_PCL}(c) shows the capacitive current $I_\mathrm{C}$ for the variations of both gate voltages at zero field. $I_\mathrm{C}$ ($\propto C_\mathrm{TG}$) shows two significant drops indicating the depletion of the PCL and the 2DES, respectively. The measured capacitance represents the sum of the two parallel capacitances related to the two plate capacitors formed by the metallic TG and the PCL and by the TG and the 2DES.
When the PCL is depleted and no longer conductive, only the capacitance between the TG and the 2DES contributes to the measured capacitance that drops as a consequence at $V_\mathrm{TG}\approx\unit{-2.5}{V}$.  The same capacitance drop as occurring at $V_\mathrm{TG}\approx\unit{-2.5}{V}$ is produced by applying $V_\mathrm{POG}\approx\unit{-2.5}{V}$ when $V_\mathrm{TG}=\unit{0}{V}$ because the PCL below the TG is separated from the drain contact where the capacitive current is measured.

Having found a method to determine the exact voltage where the PCL is depleted, we scrutinize the screening effect of the PCL from the RIs. We evaluate Sample B which has a weaker PCL, already depleted with a gate voltage of $\unit{-0.5}{V}$.
		
\begin{figure}[hbt]
	\includegraphics[scale=0.9]{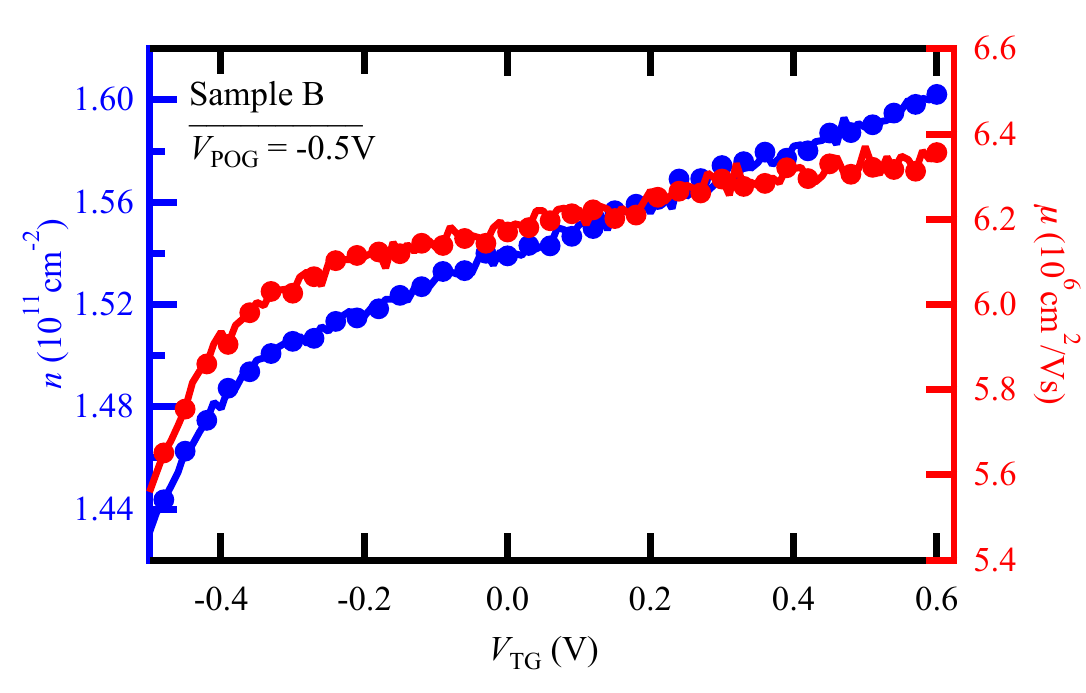}
\caption{{\sl Sample B} ($T\approx\unit{80}{mK}$): Impact of the applied top-gate voltage $V_\mathrm{TG}$ on the 2DES characteristics (electron density $n$ and mobility $\mu$).}
\label{fig:TG_dens}
\end{figure}

Figure\;\ref{fig:TG_dens} shows the electron density and mobility of the 2DES as a function of the top-gate voltage $V_\mathrm{TG}$  when $V_\mathrm{POG}=\unit{-0.5}{V}$ is applied. For \mbox{$\unit{-0.3}{V}\leq V_\mathrm{TG}\leq\unit{+0.6}{V}$}, variations in the PCL density result in a linear response of the density and mobility of the 2DES. We do not attribute this to changes in the screening of RIs but to a different effect called “compressibility” \cite{Eisenstein1994}, where the density of the 2DES is affected by density changes in the PCL. Beyond $V_\mathrm{TG}=\unit{-0.4}{V}$, the electric field from the top gate penetrates the parallel layer and can directly affect the 2DES.

\begin{figure}
		\includegraphics[scale=0.9]{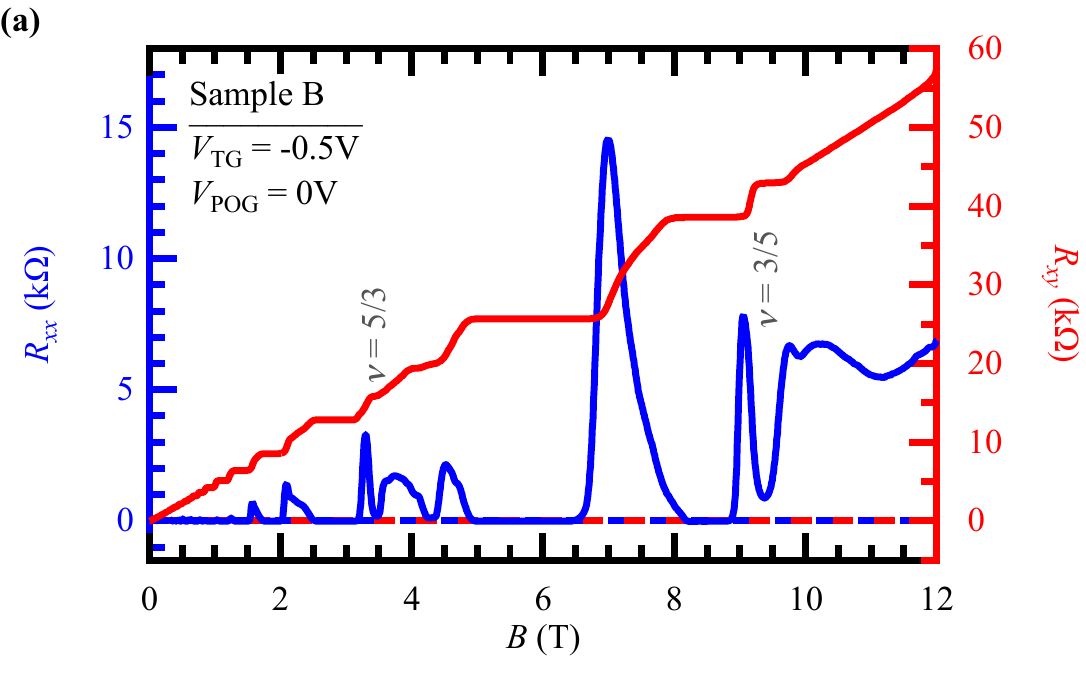}
		\includegraphics[scale=0.9]{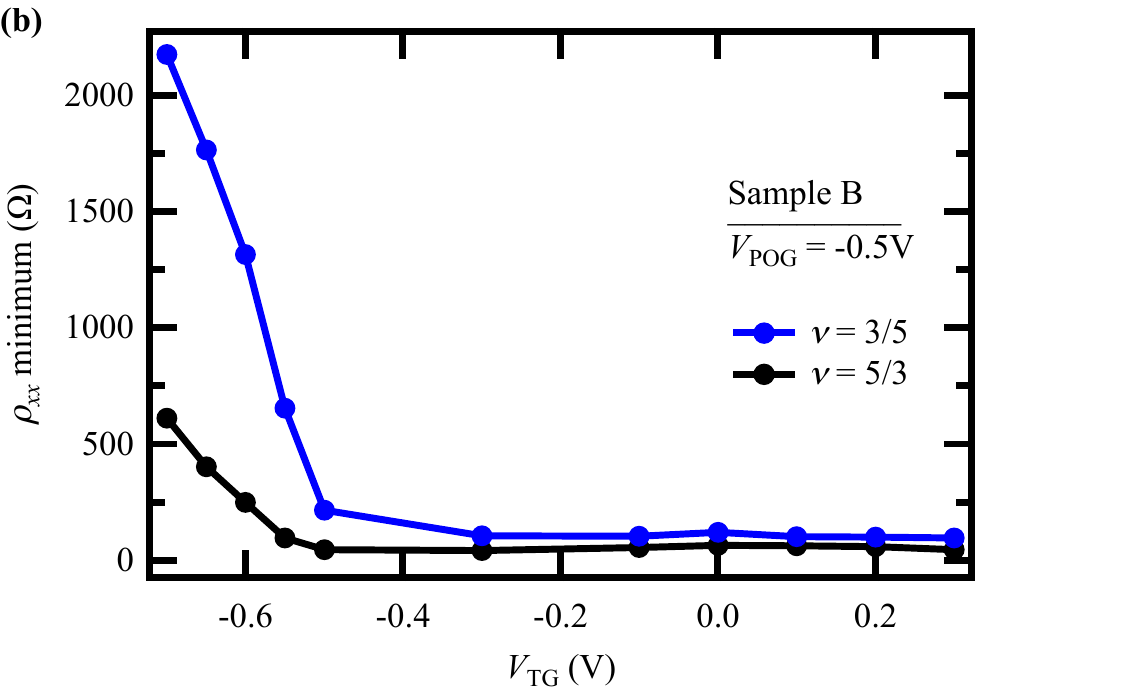}
		\includegraphics[scale=0.9]{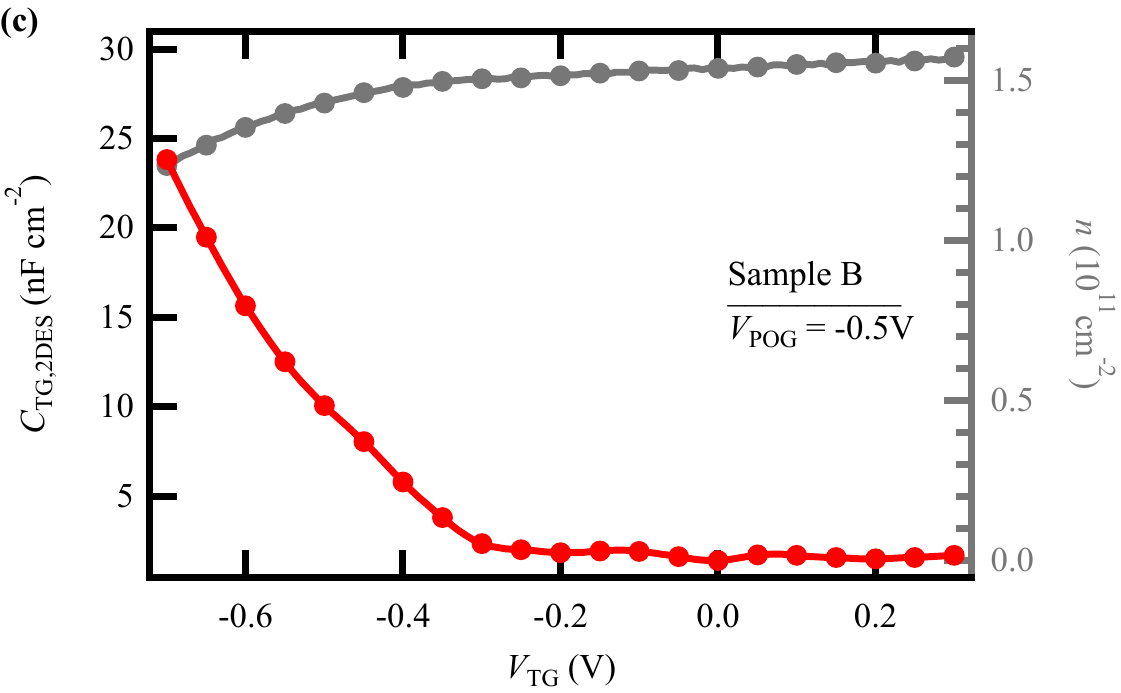}
	\caption{{\sl Sample B} ($T\approx\unit{80}{mK}$): {\bf (a)} Resistances when the parallel conducting layer is entirely depleted. {\bf (b)} Evolution of the longitudinal resistivity $\rho_{xx}$ at fractional fillings $\nu=\nicefrac{3}{5}$ (blue) and $\nu=\nicefrac{5}{3}$ (black) with the top-gate voltage $V_\mathrm{TG}$, i.e., as a function of the PCL density. {\bf (c)} The derivative of the coinciding density change (gray, right axis) with respect to $V_\mathrm{TG}$ is proportional to the capacitance per unit area $C_\mathrm{TG,2DES}$ (red).}
	\label{fig:rhoxx_min} 
\end{figure}

We deem the development of fractional quantum Hall states a better indicator of the impact of RIs\cite{Umansky2009,Reichl2014a,Peters2016} on the 2DES than the mostly background-impurity dominated electron mobility\cite{Hwang2008,Umansky1997,Reichl2014a}. The effect of the screening from the RIs by the PCL in terms of the value of the minimum in the longitudinal resistivity at two representative filling factors $\nu=\nicefrac{3}{5}$ and $\nu=\nicefrac{5}{3}$ is depicted in Fig.\;\ref{fig:rhoxx_min}. In the range of the top-gate voltage where parallel conductance occurs ($V_\mathrm{TG}>\unit{-0.5}{V}$), the minima of both FQH states are only slightly affected by $V_\mathrm{TG}$, i.e., by the density in the PCL. For $V_\mathrm{TG}<\unit{-0.5}{V}$, we observe a rapid increase in the longitudinal resistivity $\rho_{xx}$ of both states because the PCL vanishes and RIs are no longer effectively screened.

While the density of the 2DES also changes in response to the top-gate voltage, we stress that the sudden strong increase in $\rho_{xx}$ for $V_\mathrm{TG}<\unit{-0.5}{V}$ cannot be solely attributed to these comparably small changes in the electron density. We attribute the rapid variations in the longitudinal resistivity to the screening effect of the PCL.
The amount of the density change gives the capacitance per unit area according to Eq.\;\eqref{eq:CTG} via
\begin{equation}
	C_\mathrm{TG,2DES}= -e \frac{\mathrm{d}n}{\mathrm{d}V_\mathrm{TG}} \; .
\end{equation}
As shown in Fig.\;\ref{fig:rhoxx_min}(c), the value of $C_\mathrm{TG,2DES}$ does not exceed \unit{25}{nF cm^{-2}} in the investigated range of the top-gate voltage $V_\mathrm{TG}$.
The difference to the geometrical capacitance of $C_\mathrm{TG,2DES}^\mathrm{geom}\gtrsim \unit{100}{nF cm^{-2}}$ (with $C_\mathrm{TG,2DES}^\mathrm{geom} = \epsilon_0\epsilon/d$ and the distance $d=\unit{100}{nm}$ between the TG and the 2DES) reveals the existence of residual free charges in the doping layer even for $V_\mathrm{TG}\leq\unit{-0.5}{V}$. In this range, the parallel conductance of the doping layer vanishes (cf.\ Fig.\;\ref{fig:rhoxx_min}(a)). If charges in the screening layer were completely absent, the capacitance would result solely from charges in the 2DES and the top gate, i.e., the geometrical capacitance. From the tremendous increase of the resistivity minima in the range $\unit{-0.7}{V}\leq V_\mathrm{TG}\leq\unit{-0.5}{V}$, it can be deduced that even a small amount of free mobile charges in the doping layer, not forming a fully conductive layer, is sufficient to screen the RIs in a notable manner. Hence, a high doping concentration up to the threshold of parallel conductance favors the development of fractional quantum Hall states.

In summary, we have shown how a parallel-conducting doping layer can be eliminated from the magneto-transport data of a two-dimensional electron system by an appropriate gate design. The parallel conduction can be manipulated by a top gate in the Hall-bar region and we observe a very pronounced response in the development of fractional quantum Hall states which we attribute to the screening from the remote dopants. Since these effects compete with a small change in the 2DES density, for a more thorough investigation, even more sophisticated samples with top and back gates must be designed. In such samples, the density change during the variation of the PCL could be compensated by the back gate.

\begin{acknowledgments}
We would like to acknowledge Thomas Feil for helpful discussions. The financial  support by the Swiss Nation\-al Science Foundation (SNF) was highly appreciated.
\end{acknowledgments}


\begin{thebibliography}{14}%
\makeatletter
\providecommand \@ifxundefined [1]{%
 \@ifx{#1\undefined}
}%
\providecommand \@ifnum [1]{%
 \ifnum #1\expandafter \@firstoftwo
 \else \expandafter \@secondoftwo
 \fi
}%
\providecommand \@ifx [1]{%
 \ifx #1\expandafter \@firstoftwo
 \else \expandafter \@secondoftwo
 \fi
}%
\providecommand \natexlab [1]{#1}%
\providecommand \enquote  [1]{``#1''}%
\providecommand \bibnamefont  [1]{#1}%
\providecommand \bibfnamefont [1]{#1}%
\providecommand \citenamefont [1]{#1}%
\providecommand \href@noop [0]{\@secondoftwo}%
\providecommand \href [0]{\begingroup \@sanitize@url \@href}%
\providecommand \@href[1]{\@@startlink{#1}\@@href}%
\providecommand \@@href[1]{\endgroup#1\@@endlink}%
\providecommand \@sanitize@url [0]{\catcode `\\12\catcode `\$12\catcode
  `\&12\catcode `\#12\catcode `\^12\catcode `\_12\catcode `\%12\relax}%
\providecommand \@@startlink[1]{}%
\providecommand \@@endlink[0]{}%
\providecommand \url  [0]{\begingroup\@sanitize@url \@url }%
\providecommand \@url [1]{\endgroup\@href {#1}{\urlprefix }}%
\providecommand \urlprefix  [0]{URL }%
\providecommand \Eprint [0]{\href }%
\providecommand \doibase [0]{http://dx.doi.org/}%
\providecommand \selectlanguage [0]{\@gobble}%
\providecommand \bibinfo  [0]{\@secondoftwo}%
\providecommand \bibfield  [0]{\@secondoftwo}%
\providecommand \translation [1]{[#1]}%
\providecommand \BibitemOpen [0]{}%
\providecommand \bibitemStop [0]{}%
\providecommand \bibitemNoStop [0]{.\EOS\space}%
\providecommand \EOS [0]{\spacefactor3000\relax}%
\providecommand \BibitemShut  [1]{\csname bibitem#1\endcsname}%
\let\auto@bib@innerbib\@empty
\bibitem [{\citenamefont {Grayson}\ and\ \citenamefont
  {Fischer}(2005)}]{Grayson2005}%
  \BibitemOpen
  \bibfield  {author} {\bibinfo {author} {\bibfnamefont {M.}~\bibnamefont
  {Grayson}}\ and\ \bibinfo {author} {\bibfnamefont {F.}~\bibnamefont
  {Fischer}},\ }\href {\doibase 10.1063/1.1948529} {\bibfield  {journal}
  {\bibinfo  {journal} {J. Appl. Phys.}\ }\textbf {\bibinfo {volume} {98}},\
  \bibinfo {pages} {013709} (\bibinfo {year} {2005})}\BibitemShut {NoStop}%
\bibitem [{\citenamefont {Gamez}\ and\ \citenamefont
  {Muraki}(2013)}]{Gamez2013}%
  \BibitemOpen
  \bibfield  {author} {\bibinfo {author} {\bibfnamefont {G.}~\bibnamefont
  {Gamez}}\ and\ \bibinfo {author} {\bibfnamefont {K.}~\bibnamefont {Muraki}},\
  }\href {\doibase 10.1103/PhysRevB.88.075308} {\bibfield  {journal} {\bibinfo
  {journal} {Phys. Rev. B}\ }\textbf {\bibinfo {volume} {88}},\ \bibinfo
  {pages} {075308} (\bibinfo {year} {2013})}\BibitemShut {NoStop}%
\bibitem [{\citenamefont {Berl}\ \emph {et~al.}(2016)\citenamefont {Berl},
  \citenamefont {Tiemann}, \citenamefont {Dietsche}, \citenamefont {Karl},\
  and\ \citenamefont {Wegscheider}}]{Berl2016}%
  \BibitemOpen
  \bibfield  {author} {\bibinfo {author} {\bibfnamefont {M.}~\bibnamefont
  {Berl}}, \bibinfo {author} {\bibfnamefont {L.}~\bibnamefont {Tiemann}},
  \bibinfo {author} {\bibfnamefont {W.}~\bibnamefont {Dietsche}}, \bibinfo
  {author} {\bibfnamefont {H.}~\bibnamefont {Karl}}, \ and\ \bibinfo {author}
  {\bibfnamefont {W.}~\bibnamefont {Wegscheider}},\ }\href {\doibase
  10.1063/1.4945090} {\bibfield  {journal} {\bibinfo  {journal} {Appl. Phys.
  Lett.}\ }\textbf {\bibinfo {volume} {108}},\ \bibinfo {pages} {132102}
  (\bibinfo {year} {2016})}\BibitemShut {NoStop}%
\bibitem [{\citenamefont {Chambers}(1952)}]{Chambers1952}%
  \BibitemOpen
  \bibfield  {author} {\bibinfo {author} {\bibfnamefont {R.~G.}\ \bibnamefont
  {Chambers}},\ }\href {\doibase 10.1088/0370-1298/65/11/304} {\bibfield
  {journal} {\bibinfo  {journal} {Proc. Phys. Soc. Sect. A}\ }\textbf {\bibinfo
  {volume} {65}},\ \bibinfo {pages} {903} (\bibinfo {year} {1952})}\BibitemShut
  {NoStop}%
\bibitem [{\citenamefont {van Houten}\ \emph {et~al.}(1988)\citenamefont {van
  Houten}, \citenamefont {Williamson}, \citenamefont {Broekaart}, \citenamefont
  {Foxon},\ and\ \citenamefont {Harris}}]{VanHouten1988}%
  \BibitemOpen
  \bibfield  {author} {\bibinfo {author} {\bibfnamefont {H.}~\bibnamefont {van
  Houten}}, \bibinfo {author} {\bibfnamefont {J.~G.}\ \bibnamefont
  {Williamson}}, \bibinfo {author} {\bibfnamefont {M.~E.~I.}\ \bibnamefont
  {Broekaart}}, \bibinfo {author} {\bibfnamefont {C.~T.}\ \bibnamefont
  {Foxon}}, \ and\ \bibinfo {author} {\bibfnamefont {J.~J.}\ \bibnamefont
  {Harris}},\ }\href {\doibase 10.1103/PhysRevB.37.2756} {\bibfield  {journal}
  {\bibinfo  {journal} {Phys. Rev. B}\ }\textbf {\bibinfo {volume} {37}},\
  \bibinfo {pages} {2756} (\bibinfo {year} {1988})}\BibitemShut {NoStop}%
\bibitem [{\citenamefont {Kane}\ \emph {et~al.}(1985)\citenamefont {Kane},
  \citenamefont {Apsley}, \citenamefont {Anderson}, \citenamefont {Taylor},\
  and\ \citenamefont {Kerr}}]{Kane1985}%
  \BibitemOpen
  \bibfield  {author} {\bibinfo {author} {\bibfnamefont {M.~J.}\ \bibnamefont
  {Kane}}, \bibinfo {author} {\bibfnamefont {N.}~\bibnamefont {Apsley}},
  \bibinfo {author} {\bibfnamefont {D.~A.}\ \bibnamefont {Anderson}}, \bibinfo
  {author} {\bibfnamefont {L.~L.}\ \bibnamefont {Taylor}}, \ and\ \bibinfo
  {author} {\bibfnamefont {T.}~\bibnamefont {Kerr}},\ }\href {\doibase
  10.1088/0022-3719/18/29/013} {\bibfield  {journal} {\bibinfo  {journal} {J.
  Phys. C}\ }\textbf {\bibinfo {volume} {18}},\ \bibinfo {pages} {5629}
  (\bibinfo {year} {1985})}\BibitemShut {NoStop}%
\bibitem [{\citenamefont {Harris}(1991)}]{Harris1991}%
  \BibitemOpen
  \bibfield  {author} {\bibinfo {author} {\bibfnamefont {J.~J.}\ \bibnamefont
  {Harris}},\ }\href {\doibase 10.1088/0957-0233/2/12/016} {\bibfield
  {journal} {\bibinfo  {journal} {Meas. Sci. Technol.}\ }\textbf {\bibinfo
  {volume} {2}},\ \bibinfo {pages} {1201} (\bibinfo {year} {1991})}\BibitemShut
  {NoStop}%
\bibitem [{\citenamefont {Drude}(1900)}]{Drude1900a}%
  \BibitemOpen
  \bibfield  {author} {\bibinfo {author} {\bibfnamefont {P.}~\bibnamefont
  {Drude}},\ }\href {\doibase 10.1002/andp.19003060312} {\bibfield  {journal}
  {\bibinfo  {journal} {Ann. Phys.}\ }\textbf {\bibinfo {volume} {306}},\
  \bibinfo {pages} {566} (\bibinfo {year} {1900})}\BibitemShut {NoStop}%
\bibitem [{\citenamefont {Eisenstein}, \citenamefont {Pfeiffer},\ and\
  \citenamefont {West}(1994)}]{Eisenstein1994}%
  \BibitemOpen
  \bibfield  {author} {\bibinfo {author} {\bibfnamefont {J.~P.}\ \bibnamefont
  {Eisenstein}}, \bibinfo {author} {\bibfnamefont {L.~N.}\ \bibnamefont
  {Pfeiffer}}, \ and\ \bibinfo {author} {\bibfnamefont {K.~W.}\ \bibnamefont
  {West}},\ }\href {\doibase 10.1103/PhysRevB.50.1760} {\bibfield  {journal}
  {\bibinfo  {journal} {Phys. Rev. B}\ }\textbf {\bibinfo {volume} {50}},\
  \bibinfo {pages} {1760--1778} (\bibinfo {year} {1994})}\BibitemShut {NoStop}%
\bibitem [{\citenamefont {Umansky}\ \emph {et~al.}(2009)\citenamefont
  {Umansky}, \citenamefont {Heiblum}, \citenamefont {Levinson}, \citenamefont
  {Smet}, \citenamefont {N{\"{u}}bler},\ and\ \citenamefont
  {Dolev}}]{Umansky2009}%
  \BibitemOpen
  \bibfield  {author} {\bibinfo {author} {\bibfnamefont {V.}~\bibnamefont
  {Umansky}}, \bibinfo {author} {\bibfnamefont {M.}~\bibnamefont {Heiblum}},
  \bibinfo {author} {\bibfnamefont {Y.}~\bibnamefont {Levinson}}, \bibinfo
  {author} {\bibfnamefont {J.}~\bibnamefont {Smet}}, \bibinfo {author}
  {\bibfnamefont {J.}~\bibnamefont {N{\"{u}}bler}}, \ and\ \bibinfo {author}
  {\bibfnamefont {M.}~\bibnamefont {Dolev}},\ }\href {\doibase
  10.1016/j.jcrysgro.2008.09.151} {\bibfield  {journal} {\bibinfo  {journal}
  {J. Cryst. Growth}\ }\textbf {\bibinfo {volume} {311}},\ \bibinfo {pages}
  {1658} (\bibinfo {year} {2009})}\BibitemShut {NoStop}%
\bibitem [{\citenamefont {Reichl}\ \emph {et~al.}(2014)\citenamefont {Reichl},
  \citenamefont {Chen}, \citenamefont {Baer}, \citenamefont {R{\"{o}}ssler},
  \citenamefont {Ihn}, \citenamefont {Ensslin}, \citenamefont {Dietsche},\ and\
  \citenamefont {Wegscheider}}]{Reichl2014a}%
  \BibitemOpen
  \bibfield  {author} {\bibinfo {author} {\bibfnamefont {C.}~\bibnamefont
  {Reichl}}, \bibinfo {author} {\bibfnamefont {J.}~\bibnamefont {Chen}},
  \bibinfo {author} {\bibfnamefont {S.}~\bibnamefont {Baer}}, \bibinfo {author}
  {\bibfnamefont {C.}~\bibnamefont {R{\"{o}}ssler}}, \bibinfo {author}
  {\bibfnamefont {T.}~\bibnamefont {Ihn}}, \bibinfo {author} {\bibfnamefont
  {K.}~\bibnamefont {Ensslin}}, \bibinfo {author} {\bibfnamefont
  {W.}~\bibnamefont {Dietsche}}, \ and\ \bibinfo {author} {\bibfnamefont
  {W.}~\bibnamefont {Wegscheider}},\ }\href {\doibase
  10.1088/1367-2630/16/2/023014} {\bibfield  {journal} {\bibinfo  {journal}
  {New J. Phys.}\ }\textbf {\bibinfo {volume} {16}},\ \bibinfo {pages} {023014}
  (\bibinfo {year} {2014})}\BibitemShut {NoStop}%
\bibitem [{\citenamefont {Peters}\ \emph {et~al.}(2016)\citenamefont {Peters},
  \citenamefont {Tiemann}, \citenamefont {Reichl},\ and\ \citenamefont
  {Wegscheider}}]{Peters2016}%
  \BibitemOpen
  \bibfield  {author} {\bibinfo {author} {\bibfnamefont {S.}~\bibnamefont
  {Peters}}, \bibinfo {author} {\bibfnamefont {L.}~\bibnamefont {Tiemann}},
  \bibinfo {author} {\bibfnamefont {C.}~\bibnamefont {Reichl}}, \ and\ \bibinfo
  {author} {\bibfnamefont {W.}~\bibnamefont {Wegscheider}},\ }\href {\doibase
  10.1103/PhysRevB.94.045304} {\bibfield  {journal} {\bibinfo  {journal} {Phys.
  Rev. B}\ }\textbf {\bibinfo {volume} {94}},\ \bibinfo {pages} {045304}
  (\bibinfo {year} {2016})}\BibitemShut {NoStop}%
\bibitem [{\citenamefont {Hwang}\ and\ \citenamefont {{Das
  Sarma}}(2008)}]{Hwang2008}%
  \BibitemOpen
  \bibfield  {author} {\bibinfo {author} {\bibfnamefont {E.~H.}\ \bibnamefont
  {Hwang}}\ and\ \bibinfo {author} {\bibfnamefont {S.}~\bibnamefont {{Das
  Sarma}}},\ }\href {\doibase 10.1103/PhysRevB.77.235437} {\bibfield  {journal}
  {\bibinfo  {journal} {Phys. Rev. B}\ }\textbf {\bibinfo {volume} {77}},\
  \bibinfo {pages} {235437} (\bibinfo {year} {2008})}\BibitemShut {NoStop}%
\bibitem [{\citenamefont {Umansky}, \citenamefont {De-Picciotto},\ and\
  \citenamefont {Heiblum}(1997)}]{Umansky1997}%
  \BibitemOpen
  \bibfield  {author} {\bibinfo {author} {\bibfnamefont {V.}~\bibnamefont
  {Umansky}}, \bibinfo {author} {\bibfnamefont {R.}~\bibnamefont
  {De-Picciotto}}, \ and\ \bibinfo {author} {\bibfnamefont {M.}~\bibnamefont
  {Heiblum}},\ }\href {\doibase 10.1063/1.119829} {\bibfield  {journal}
  {\bibinfo  {journal} {Appl. Phys. Lett.}\ }\textbf {\bibinfo {volume} {71}},\
  \bibinfo {pages} {683} (\bibinfo {year} {1997})}\BibitemShut {NoStop}%
\end{thebibliography}
%

\end{document}